# A health telemonitoring platform based on data integration from different sources

Gianluigi Ciocca, Paolo Napoletano, Matteo Romanato, Raimondo Schettini
*Department of Informatics, Systems, and Communication*
*University of Milano-Bicocca, Viale Sarca 336, 20126, Milan, Italy*
{gianluigi.ciocca, paolo.napoletano, raimondo.schettini}@unimib.it, {m.romanato}@campus.unimib.it

*Abstract*—The management of people with long-term or chronic illness is one of the biggest challenges for national health systems. In fact, these diseases are among the leading causes of hospitalization, especially for the elderly, and huge amount of resources required to monitor them leads to problems with sustainability of the healthcare systems. The increasing diffusion of portable devices and new connectivity technologies allows the implementation of telemonitoring system capable of providing support to health care providers and lighten the burden on hospitals and clinics. In this paper, we present the implementation of a telemonitoring platform for healthcare, designed to capture several types of physiological health parameters from different consumer mobile and custom devices. Consumer medical devices can be integrated into the platform via the Google Fit ecosystem that supports hundreds of devices, while custom devices can directly interact with the platform with standard communication protocols. The platform is designed to process the acquired data using machine learning algorithms, and to provide patients and physicians the physiological health parameters with a user-friendly, comprehensive, and easy to understand dashboard which monitors the parameters through time. Preliminary usability tests show a good user satisfaction in terms of functionality and usefulness.

## I. INTRODUCTION AND BACKGROUND

Nowadays, the digital revolution, with its many applications and devices, permeates every aspect of our society, causing radical transformations in many fields, and in the medical profession as well. In particular, more and more sensors and Internet of Things (IoT) devices are being used to monitor vital parameters. This is very important for patients with chronic conditions that need to be continuously monitored. Chronic conditions are a major cause of hospitalization, especially for the elderly, and are still too often treated through manual monitoring that can be prone to procedural errors and limited to outpatient visits [1].

The increase in life expectancy triggers the growth in the number of people with these conditions and to the inevitable physiological decline past a certain age, both in cognitive and physical abilities, reducing the person's independence and increasing the need for care, thus causing congestion within health care facilities.

[1]https://www.who.int/news-room/fact-sheets/detail/noncommunicable-diseases

Modern technologies can reduce the need for hospitalization in favor of a remote monitoring, for example at home [1]. In fact, public health agencies are mobilizing to offer telemonitoring services in order to provide the necessary care to as many people as possible remotely, avoiding continuous movement of patients to clinics.

The benefits provided by telemonitoring services are numerous in several aspects. From the point of view of the patients, they feels more autonomous in managing their condition and confident that they are being followed even from home. Moreover they are stimulated in complying with therapy and adopting a correct lifestyle. From the point of view of the health workers, they can work in smart working or in the hospital and, at the same time, monitor more patients while always offering quality care. Moreover, the National Health System also benefits from a reduced number of emergency room admissions, avoiding possible overload of facilities [2].

Telemonitoring systems and applications proposed in the literature offer different types of solutions: focused on telemonitoring chronic patients [3]; focused on telemonitoring acute conditions [4], [5]; focused on supporting patients remotely so as not to overcrowd hospital wards [6], [7]; focused on implementing smarthome for elderly care [8], [9].

Table I summarizes available systems and applications developed for commercial or research purposes. The table reports the references, a description of the use cases, the vital parameters that are monitored, the enabling technologies used and the type of developed system. Three of the reported systems are commercial ones while the remaining are systems developed by research groups. Some of these systems are designed for specific health conditions such as stroke prevention. The majority of the systems are designed to monitor a wide range of vital parameters. With respect to the technologies used, to communicate with the sensors and devices, most of the systems rely on standard messaging protocols for the Internet of Things (IoT) such as MQTT (Message Queue Telemetry Transport) or Bluetooth ones (BLT). Almost every systems is designed with a client-server or cloud architecture with smartphone or web-based front-end applications.

One issue that has emerged from the articles in the state of the art is the integration of systems with sensors or medical devices already on the market. Many telemonitoring systems work on custom devices suitable for its purposes without considering the devices that are already commercially available and that can monitor diverse vital parameters. Only the work by Sanna et al. [9] provides a solution that feature the use of devices from different manufacturers that work in synergy for patient care.

TABLE I
EXAMPLES OF EXISTING TELEMONITORING SYSTEMS AND APPLICATIONS.

| Reference | Use case | Vital parameters | Enabling technologies | Type | Year |
|---|---|---|---|---|---|
| Winpack [10] | Modular platform designed for continuous, real-time monitoring of the most important vital parameters | Heart rate, oxygenation, body temperature, blood pressure, blood sugar, body weight, physical activity | Proprietary | Commercial | 2017 |
| Telbios [11] | System designed for home monitoring | Heart rate, oxygenation, blood pressure, blood sugar, body weight, physical activity | Web application | Commercial | 2020 |
| Emilia-Romagna regional project [12] | System designed for health monitoring at home to ensure care in rural centers | Heart rate, oxygenation, blood pressure, body weight, physical activity | Smartphone, sensors | Commercial | 2021 |
| Wearable Sensing and Telehealth Technology with Potential Applications in the Coronavirus Pandemic [7] | Telemonitoring for patients at risk of heart attack during pandemic | Heart rate, oxygenation, blood pressure, body temperature, respiratory rate, coughing | Not specified | Research | 2020 |
| Smart home technologies for telemedicine and emergency management [8] | Telemonitoring system for elderly people through a smarthome | Heart rate, blood pressure, physical activity | Smartphone, sensors, BLT, client-server architecture | Research | 2013 |
| A Telemedicine Service System Exploiting BT/BLE Wireless Sensors for Remote Management of Chronic Patients [3] | Telemonitoring system for chronic patients based on bluetooth devices | Heart rate, oxygenation, blood pressure, respiratory rate, blood sugar, body weight | Sensors, BLT, Client-server architecture | Research | 2019 |
| Home Telemonitoring for vital signs using IoT technology [6] | Telemonitoring system for neurodegenerative diseases with real-time collection of vital signs | Heart rate, oxygenation, body temperature, blood pressure, blood sugar, body weight, physical activity | Sensors, MQTT, cloud-based arhitecture, mobile application | Research | 2016 |
| Remote web based ECG monitoring using MQTT protocol for IoT in healthcare [4] | Heartbeat-centered telemonitoring | ECG | Sensors, Raspberry, MQTT, cloud-based architecture | Research | 2018 |
| Monitor human vital signs based on IoT technology using MQTT protocol [5] | Telemonitoring system allowing real time collection of vital and environmental parameters | Heart rate, oxygenation, body temperature | Sensors, MQTT, cloud-based architecture | Research | 2020 |
| Smart Healthcare Monitoring System Using MQTT protocol [13] | Telemonitoring system for chronic illness | Heart rate, oxygenation, body temperature, acceleration, blood glucose | Sensors, Arduino, MQTT, cloud-based architecture | Research | 2018 |
| Analysis of a Telemonitoring System based on a Bluetooth Body Area Network using Smartphones [14] | General patient health telemonitoring system | Heart rate, oxygenation | Sensors, smartphone, BLT, client-server architecture | Research | 2011 |
| Iot platform for ageing society: the smart bear project [9] | IoT platform for telemonitoring of the elderly | Heart rate, oxygenation, body temperature, physical activity, body weight, hearing | Commercial medical devices, BLT, smartphones | Research | 2021 |

In this paper we propose the design of a telemonitoring platform able to collect physiological health parameters from different sources, both consumers and customs, manage the physician-patient relationship, analyze and display the data on a user-friendly, Web-based, application. The platform offers a technological solution that enables remote monitoring of patients' health status through networked devices and systems for sharing the measured clinical parameters. Parameters measured by the patient in full autonomy through wearable devices, are automatically transmitted to the platform and shared with the care team in real time, allowing the evolution of the pathology and people's health to be constantly monitored.

## II. PROPOSED PLATFORM

In this section we describe the design of our telemonitoring platform. The aims of the platform is to allow users to collect data regarding their health that come from different sources on a single web-based application. The data can be stored and, with the help of advanced AI solutions, automatically analyzed to infer potential health issues that can trigger warnings or alarms. The raw and processed data is also shared with the user's doctor that can monitor the recorded vital signals, and promptly act by devising the proper treatments and follow up actions if necessary.

### A. Overall Architecture

The high-level schema of the proposed health telemonitoring platform il illustrated in Figure 1. The platform backend is composed by four macro-blocks: data acquisition, data processing, data management, and data visualization.

Data acquisition comprises the sensor devices, and procedures to automatically collected values to be stored in the telemonitoring platform. Data processing refers to processing and machine learning algorithms that can be applied to the collected values for visualization or inference purposes. Data management is responsible for the user/system interaction and safe storage/access to the data. Finally, data visualization refers to the modules and techniques used to presents in a compact but understandable way the raw and processed data to the users and medics. The front-end part of the platform is a Web-based application used to interact with the system and monitor the collected data by the patients, caregivers and physicians enrolled in the platform.

There are a large number of commercially available medical devices, with new ones being proposed every day and increasingly connected to our smartphones and our lives. One of the aims of the project is to take advantage of consumer devices and sensors already available on the market and thus to integrate them into the proposed telemonitoring platform.

To integrate the sensor in the platform, suitable APIs have been developed. Manufacturers such as Xiaomi, Huawei, and Samsung made available their Application Programming Interfaces (APIs) to be used in custom applications [15]–[17]. For example, Samsung APIs offer the possibility of covering a good chunk of the wearable device market, however, the number of physiological parameters is quite limited. Similarly for the other competitors. Moreover, using sensors exclusively from a single manufacturer may compromise the scalability and usability of the platform.

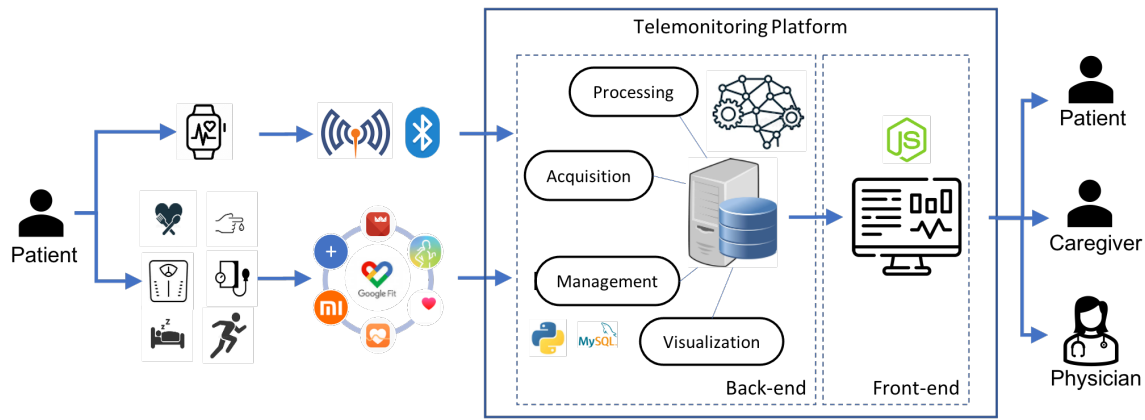

Fig. 1. Overview of our health telemonitoring platform.

For these reasons, we decide to rely on the Google Fit service [18]. Google introduced Google Fit in 2015 and has been constantly updating it also in collaboration with the World Health Organization (WHO) and the American Heart Association, with the goal of encouraging users to engage in daily physical activity by taking advantage of the various sensors on board the smartphone. Google Fit with one of its latest updates introduced a new functionality that allows personal data to be accessed from all connected apps and devices. We exploited this functionality to incorporate different data sources in our platform. All data available on Google Fit is categorized into one of the following groups:

- *activity*: daily goals, weekly goal, calories, heart points, steps, distance, minutes of movement, step cadence, cycling cadence, wheel speed, speed, power;
- *body measurements*: weight, body fat, height;
- *vital parameters*: respiratory rate, heart rate, resting heart rate, blood pressure, blood sugar, oxygen saturation, body temperature;
- *nutrition*: calorie consumption, hydration;
- *sleep*: sleep duration divided into stages;
- *cycle monitoring*: menstrual period.

The use of Google Fit requires that the patient has a compatible account. To avoid possible security and privacy issues, the hospital or institution that will provide the telemonitoring service should provide to each individual patient a personal account. This will avoid the use of personal user accounts that might contain private information irrelevant to the purpose of telemonitoring. Mobile applications supporting the Google Fit platform can synchronize their data on the cloud. The stored data is then collected by the telemonitoring platform via the implemented APIs.

To leverage the Google Fit service within our telemonitoring platform we registered our web service in the Google Cloud Platform and activate the Google Fit API. Once activated, the provided credentials, a Client ID and a Client secret, will be used in REST calls. Moreover, the list of allowed redirection URIs needs to be setup. This list is used in the user authentication phase, where it is necessary to log in with a Google profile (the one provided by the hospital or doctor) to allow the required data collection.

*B. Consumer applications*

We experiment the data acquisition process using two common apps: FatSecret [2] and MedM Health [3]. The first is a food diary app with a very large database of food products with their nutritional values that can be used to collect important data for monitoring a patient's diet. The second app supports more than 550 medical devices, including: activity tracker, glucometer, electronic sphygmomanometer, scales, saturimeter, sleep tracker, thermometer, and many others.

We designed our platform following a RESTful web service architecture. In order to collect the data from the Google Fit ecosystem, the telemonitoring platform leverage a *request and response* system that interacts with the APIs. To start the data acquisition, it is first necessary to configure the Google Fit application on the smartphone with the account given by the hospital or institution, and enabling the synchronization with third-party apps. After all the mobile applications are configured, data are automatically collected as soon as a measurement is taken through the configured device. To obtain the collected values an access token, provided after proper authentication to Google, is stored in an encrypted session variable and is included in the header of each request. This ensure security and avoids performing an authentication step with each request.

*C. Custom device*

Although we have have mainly leveraged the Google Fit service, our platform is able to integrate custom devices via standard communication protocol as well such as MQTT or BLT. In fact, for testing this functionality, we have designed a smart bracelet using consumer components that is able to collect vital signs from the user and send them via MQTT messaging, and successfully integrated it in our telemonitoring platform. A sketch of the smart bracelet is showed in Fig. It is based on the Arduino MKR 1000 board and equipped with the following sensors:

- Finger-clip Heart rate sensor for monitoring heart beat;
- GSR sensor for monitoring the electrodermal actitvity;
- WaveShare 10 IMU Sensor for monitoring the activity;
- Temperature sensor for monitoring the body temperature.

[2]https://www.fatsecret.com/
[3]https://www.medm.com/apps/health/

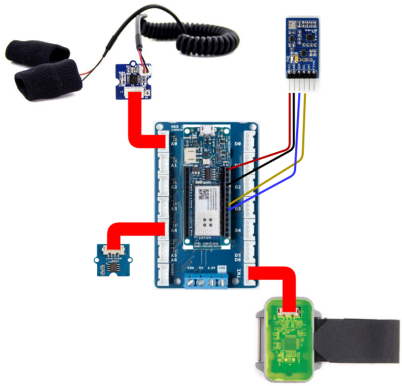

Fig. 2. Smart bracelet based on Arduino MKR 1000

The Arduino board has a wireless connection and communicates with the platform through the MQTT protocol by using the *publish* and *subscribe* strategies provided by the protocol.

### D. Developed platform

As introduced in the previous Section, our telemonitoring platform is Web-based and, in addition to managing the data collection and processing phases, is responsible for providing an interface for parameter visualization and facilitating physician-patient/caregiver interaction.

The front-end of the platform is a Web application devoted to present in an accessible way the collected data and information extracted from it. Data visualization is of paramount importance to the system; it makes it possible for a physician to visualize the collected data through visual elements such as diagrams and graphs, which create an accessible and a quick solution for observing and understanding trends, outliers, and recurrences. The front-end of the system consists of dynamic HTML pages that leverage the Bootstrap[4] library for the graphical part, and Javascript codes for graph creation.

The back end of the system is responsible for the collection, analysis and storage of the data, as well as as managing the requests via the APIs to and from the web application and the medical apps. The back-end is implemented with Node.Js[5] integrated with the Express framework[6] for route management and APIs. In addition to the graphical aspects, the platform provides various features in the areas of security, usability, and management of the system in its entirety.

To support these activities, additional modules are used:

- *express-session*[7]: to ensure the creation of user sessions;
- *https*: to ensure a secure data transmission protocol, through a key and certificate obtained via OPENSSL;
- *crypto-js*[8]: for encryption of sensitive data in the database;
- *mysql2*[9]: for creating a connection to the database used.

Finally, the middelware *passport*[10] for Node.js allow for secure authentication and registration.

The registration step is different for the patient and the doctor. First, when the patient agrees to utilize the telemonitoring platform, the hospital or the doctor provide him with the google account to be used with the system. Then, during registration the patient must fill a form with his account information: name, social security number, email and password. Next, he is prompted insert the ID of the custom devices to add to the system (e.g. those provided by the hospital). Finally, the last task to be performed is the association with the doctor that the patient agrees to share his data with. For a doctor, the registration consists of a single step that requires to enter the doctor's personal information including the medical specialization, and password.

The authentication step for the patient consists in using the credentials selected in the registration step, and then the patient must select the Google account assigned by the hospital to be redirected to his private area. A doctor first uses his credentials and then she/he selects the patient to be monitored from the list of registered patients. To avoid usability problems dictated by distraction, the patients and doctors pages of the web application have different color schemes to hint the users if they are in the correct section.

Figure 3 shows a user's dashboard that can be accessed from the web application. The dashboard allows the patient and doctors to monitor instant values of vital parameters as well as collected time series. Analysis of the data, when it is available, is also reported. Each parameter is shown in a separate card for ease of readability. Other pages in the web application are related to the management of the devices and account and are not shown here.

### III. USABILITY

To evaluate usability aspects of the the telemonitoring platform, we designed a test and we involved five users. The test consists of two parts: execution of several tasks and answering of a questionnaire. The execution of the tasks is aimed at evaluating the proper functionality of the platform and the correct integration of the Google Fit service within the platform as described in Section II. At the end of the first part of the test, participants are asked to answer a questionnaire referring to the system, with the aim of evaluating its effectiveness, efficiency, and user satisfaction.

---

[4] https://getbootstrap.com/
[5] https://nodejs.org/it/
[6] https://expressjs.com/it/21

[7] https://www.npmjs.com/package/express-session
[8] https://www.npmjs.com/package/crypto-js
[9] https://www.npmjs.com/package/mysql2
[10] https://www.passportjs.org/

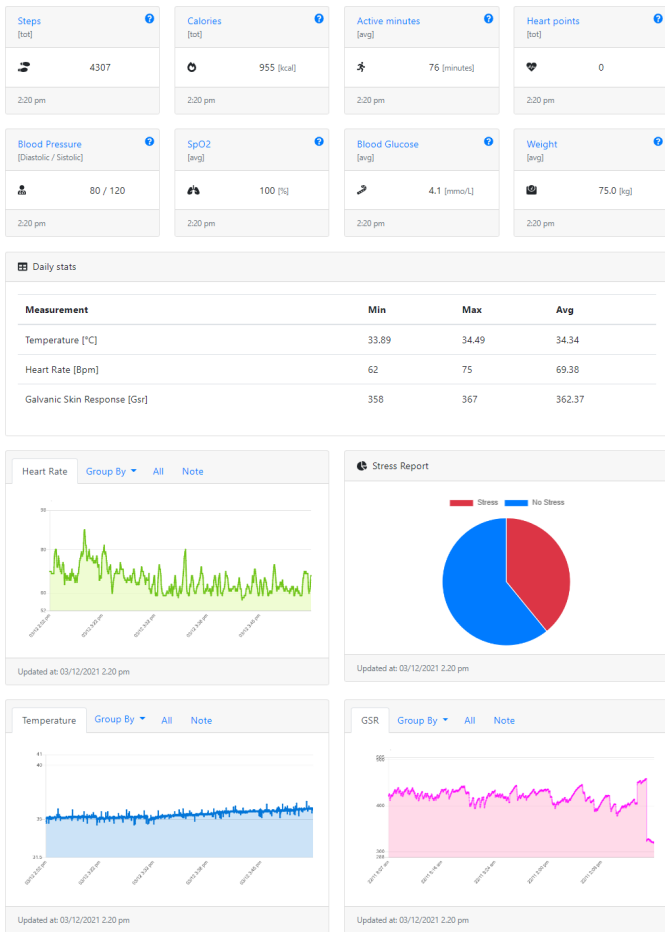

Fig. 3. Sample of a patient's dashboard with the vital parameters monitored and analyzed.

TABLE II
RESULTS OF THE TASK PERFORMED BY THE USERS. M<30: MALE UNDER 30 YEARS OLD; F: FEMALE. ✓: COMPLETED WITHOUT HELP; -: COMPLETED WITH A SMALL HINT; X: COMPLETED WITH HELP.

| Task | User 1 M<30 | User 2 M<30 | User 3 30≤F≤60 | User 4 30<M<60 | User 5 F>60 |
|---|---|---|---|---|---|
| 1 | ✓ | ✓ | - | - | - |
| 2 | ✓ | ✓ | ✓ | ✓ | - |
| 3 | ✓ | ✓ | - | - | - |
| 4 | ✓ | ✓ | - | ✓ | ✓ |
| 5 | ✓ | ✓ | ✓ | ✓ | ✓ |
| 6 | ✓ | ✓ | ✓ | - | - |
| 7 | ✓ | ✓ | ✓ | ✓ | ✓ |
| 8 | ✓ | ✓ | ✓ | ✓ | ✓ |
| 9 | ✓ | ✓ | ✓ | ✓ | ✓ |
| 10 | ✓ | ✓ | ✓ | ✓ | ✓ |

The first part of the test consists of the execution of several tasks:

1) Register in the platform;
2) Disconnect from the portal;
3) View the values of sleep performed in the last month;
4) View last week's pressure and add a note on the highest value;
5) Measure the pressure through the sensor and view the daily graph;
6) Add a new doctor to the assigned doctors;
7) Remove an assigned doctor;
8) Change the name of the device;
9) Add a new device;
10) Edit a field of your choice in the health profile.

Testers are given one minute to familiarize themselves with the interface, and then tasks are administered in the given order. Each task can be passed in three ways: Without any help, with a small hint and by help. In Table II we report the results of the tasks performed by the five users. All the user are able to complete the task autonomously or with a small hint. One issue that arose during the tasks is the English language constraint, which limited the understanding of navigating between web pages by sometimes requiring small hints.

The second part of the test consists of the administration of a questionnaire consisting of 15 statements. The first 10 taken from the SUS (System Usability Scale) questionnaire developed by John Brooke [19], while the remaining five are specifically designed to assess aspects peculiar to the proposed platform. Responses to each field are given through a range from 1 (strongly disagree) to 5 (strongly agree). The purpose of the questionnaire is to obtain an overall evaluation of the system through usability, usefulness, and trustworthiness. The last question is designed from a marketing perspective. Between the advertisement of a product, and the advice of a person we know, we always tend to trust more the acquaintance.

Table III shows the statements of the questionnaire and the corresponding average responses, which are particularly concordant since the mean value of the standard deviation is less than one. Please note that in the top 10 statements, high values in the odd statements are the best results, while for even statements lower values are best. From the average results, it is evident a general positive sentiment dictated by the ease of use of the platform to the extent that the users would recommend it to acquaintances. No particular issues have been reported apart for a multi-language translation of the user interface.

IV. CONCLUSION

The proposed telemonitoring platform represents a solution that can acquire and process data from different devices, both consumer and custom, and that can provide visualization of monitored physiological parameters through graphs. The goal of this solution is to provide patients with continuous, quality care by automating home monitoring, supporting the physician's work so that he has a complete and accurate view of the patient's situation, and preventing unnecessary admissions to hospital facilities.

The telemonitoring platform consists of three main steps: the collection of data through sensors, the processing of some of these to make estimates of the patient's physical condition, and the visualization of the monitored parameters. Data acquisition can be accomplished through custom and consumer medical devices that are integrated in the system via standard communication protocols or the Google Fit ecosystem. This allows the system to leverage a large portion of the wearable market through the various available applications. The system allows the inclusion of machine learning algorithms to support the automatic analysis of the collected data and infer possible patient's conditions or sickness. Finally, the front-end of the system is designed with an intuitive interface to provide the users with the necessary tools to analyze and monitor the health of the patients.

TABLE III
SYSTEM USABILITY SCORE RESULTS. WE REPORT THE AVERAGE AND STANDARD DEVIATIONS OF THE SCORES (FROM 1 TO 5). SEE TEXT ON HOW TO INTERPRET THE VALUES FOR THE DIFFERENT STATEMENTS.

| # | Statements | Average | Dev.Std |
|---|---|---|---|
| 1 | I think that I would like to use this system frequently | 3.8 | 0.45 |
| 2 | I found the system unnecessarily complex | 1.6 | 0.89 |
| 3 | I Though the system was easy to use | 4.2 | 1.10 |
| 4 | I think that I would need the support of a technical person to be able to use this system | 2.4 | 1.14 |
| 5 | I found the various functions in this system were well integrated | 4.6 | 0.55 |
| 6 | I thought there was too much inconsistency in this system | 1.8 | 0.84 |
| 7 | I would imagine that most people would learn to use this system very quickly | 3.8 | 0.45 |
| 8 | I found the system very cumbersome to use | 4.0 | 0.00 |
| 9 | I felt very confident using the system | 3.8 | 1.30 |
| 10 | I needed to learn a lot of things before I could get going with this system | 2.2 | 1.64 |
| 11 | I think the data collected is accurate | 4.2 | 0.45 |
| 12 | I think the stress detected is accurate | 4.0 | 0.71 |
| 13 | I think the graphs used make it easier to understand the data | 5.0 | 0.00 |
| 14 | I think the system strengthens the relationship between doctor and patient | 3.4 | 0.89 |
| 15 | I would recommend it to my acquaintances | 4.4 | 0.55 |

To test the usefulness of the proposed system, we performed usability tests with different users. Results show that the system could be further improved in several ways.The front-end application needs multilingual support. For the back-end, we plan to incorporate and assess several analysis tools aimed at detecting possible health issues in a preemptive way. Acceptability assessment is also needed. We plan to perform such analysis with users and physicians. Finally, environmental conditions could influence the wellness of a patient. For this reason, we plan to investigate if ambient sensors could be exploited and added to the system to understand which environmental factors-such as cold, noise, air quality-could may affect, positively or negatively, the patient's specific pathology.